\documentclass[9pt,a4paper,twocolumn]{amsart}
\usepackage[T1]{fontenc}
\usepackage{url}
\usepackage[slantedGreek]{cmbright}
\usepackage[mathcal]{euscript}
\usepackage{latexsym}
\usepackage{amssymb}
\usepackage{amsmath}
\usepackage{amsfonts}
\usepackage{extarrows}
\usepackage{graphics,graphicx}
\usepackage{pst-all}
\usepackage{esrelation}
\usepackage{nonfloat}
\usepackage{cite}
\usepackage[]{hyperref}
\hypersetup{colorlinks=true}
\usepackage{amsrefs}%
\usepackage[yyyymmdd,hhmmss]{datetime}
\usepackage{anysize}
\marginsize{12mm}{12mm}{12mm}{12mm}

\def\bean{\begin{equation*}\begin{aligned}}
\def\eean{\end{aligned}\end{equation*}}

\def\bea{\begin{equation}\begin{aligned}}
\def\eea{\end{aligned}\end{equation}}

\def\ba{\begin{align*}}
\def\ea{\end{align*}}

\def\bc{\begin{center}}
\def\ec{\end{center}}

\def\bfi{\begin{figure}}
\def\efi{\end{figure}}

\def\bp{\begin{pspicture}}
\def\ep{\end{pspicture}}

\def\p{\partial}

\def\E{\mathcal{E}}

\def\d{\mathrm{d}}
\def\e{\varepsilon}
\def\f{\varphi}

\def\r{\rho}

\definecolor{vert1}{rgb}{0,.6,.3}
\definecolor{vert}{rgb}{0,.5,.15}
\definecolor{VertClair}{rgb}{.16,1,.36}
\definecolor{violet}{rgb}{1,.6,1}
\definecolor{beige}{rgb}{1,.85,.65}
\definecolor{Azur}{rgb}{.3,.9,1}
\definecolor{rose}{rgb}{1,.85,.95}
\definecolor{Orange}{rgb}{1,.6,.1}
\definecolor{Jaune}{rgb}{1,.84,.24}
\definecolor{gris90}{rgb}{.9,.9,.9}
\definecolor{gris95}{rgb}{.95,.95,.95}
\definecolor{gris80}{rgb}{.8,.8,.8}

\makeatletter
\g@addto@macro{\endabstract}{\@setabstract}
\newcommand{\authorfootnotes}{\renewcommand\thefootnote{\@fnsymbol\c@footnote}}%
\makeatother

\date{This manuscript was compiled on {\today} at {\currenttime}.}

\begin{document}

\title{Topology of Protocells : do nanoholes catalyse fission ?}
\author{Romain ATTAL}

\maketitle

\begin{abstract}
We propose a mechanism with a low activation energy for lipid translocation, 
based on a change of topology of the membrane of a protocell. 
The inner and outer layers are connected and form toroidal nanoholes 
stabilised by repulsive electrostatic forces for small radius and
attractive elastic forces for large radius. Thanks to these holes, 
the energy barrier of translocation is drastically reduced and 
a difference of temperature between the inside and the outside 
of the protocell can induce a differential growth of these layers, 
until the vesicle splits in two.
\end{abstract}




\section{\bf Introduction}
One of the enigmas of the origin of life is to find out a plausible 
sequence of processes which pushes abiotic matter to form autonomous,
self-reproducing protocells \cite{Pro,Mor3}. We leave aside the problem
of the formation of adequate amphiphilic molecules in the primordial 
soup or in deep hydrothermal systems and we start from the hypothesis 
that cellularity is at the basis of life \cite{Mor2}. 
Being endowed with the metabolic ability to split, 
our \emph{First Unicellular Common Ancestors} (FUCA) could initiate 
a Darwinian selection process, after a sufficiently large
landscape of chemical reactions has been explored to select reproducible,
auto-catalytic metabolisms. This prebiomass would then be 
able to grow exponentially in the marine cradle of Life. 
Part of the problem is thus to find a plausible, non-anachronistic 
mechanism for self-reproduction of protocells, without 
complex molecules which result from a long biochemical
evolution. Our fundamental hypothesis is that the ability to split came 
so early (around four billion years ago) that it must be based on 
simple physical forces. 
In \cite{TDFP}, we proposed that a difference of temperature between
the inside and the outside of protocells could have triggered their
fission. The hottest molecules of the inner layer could jump to the 
outer layer where they would thermalise with the colder environment.
The membrane then acts as a soft, closed Knudsen barrier \cite{Mor1}.
The outer layer growing more quickly than the inner layer, the total mean
curvature of the median surface increases steadily until the membrane splits.
However, the activation energy of this translocation (flip-flop) 
process is quite high. How can we lower this energy barrier ?
In the present work, we propose that the formation of stable toroidal 
holes connecting the two layers of the membrane could help their
molecules to bypass this obstacle.

Our article is organised as follows.
In section 2, we give a short description of our model 
of thermally driven fission of protocells \cite{TDFP}.
In section 3, we pierce the membrane and prove the (meta)stability
of nanoholes. 
In section 4, we compute the molecular flux through a single hole
induced by the temperature gap between the inside and the outside.
We conclude with the problem of the structure of water at this
nanometric scale. Does it further stabilise the nanoholes of the
membrane ?

\section{\bf Thermally driven fission of protocells}

In this section, we recall the main ideas, hypotheses and results 
of the model we developed in \cite{TDFP}.
We suppose that protocells are mere vesicles encapsulating some
exothermic metabolism and that the membrane molecules are synthesised
inside. The inner temperature $(T_1)$ is larger than the outer
temperature $(T_0)$. This temperature gap is maintained in a steady
state : the nutrient molecules, present in the surroundings in
constant concentrations, enter the vesicle. The waste molecules 
are expelled to the outside, where their concentration is lower
than inside.
Following the laws of non-equilibrium thermodynamics \cite{KC},
we can write down differential equations for the growth of the areas,
$A_0$ and $A_1$, of the leaflets of the membrane. 
In a steady state, both grow exponentially but at different rates. 
This differential growth makes the total mean curvature of the median
surface, $\iint H \propto A_0-A_1$, grow also exponentially.
If the cell is initially cylindrical, a small reduction of its
radius increases until the radius goes to zero and the protocell
splits in two.

This model is based on the ability of membrane molecules to
go from the inner leaflet to the outer leaflet, without the
help of dedicated membrane proteins like flippase and floppase  \cite{Mou}.
The energy cost of this translocation is however very high, 
between $50$ and $100$ kJ/mol, according to \cite{JC}. Therefore, 
It seems necessary to bypass this barrier by avoiding the energy cost 
of a hydrophilic polar head surrounded by hydrophobic aliphatic chains. 
This can be achieved by piercing the membrane and letting the molecules 
of the inner leaflet flow through the holes to join the outer leaflet,
as we will see in the next section.

\section{\bf Hole stability}

\begin{figure}
\centering
\includegraphics[width=.9\linewidth]{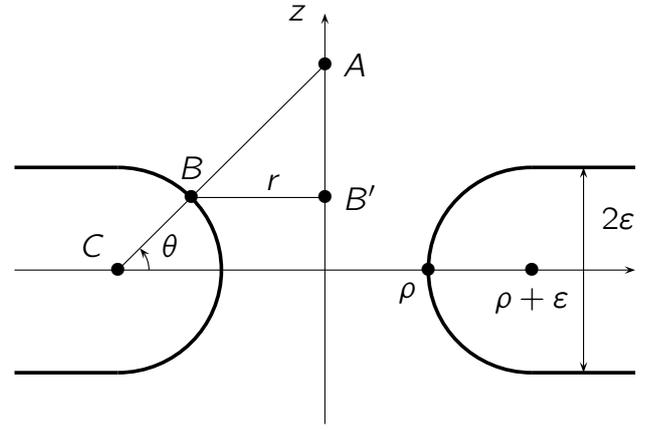}
\figcaption{Meridian section of a hole of radius $\r$, 
in a bilayer of thickness $2\e$.}
\label{fig:hole1}
\end{figure}

\begin{figure}
\centering
\includegraphics[width=.9\linewidth]{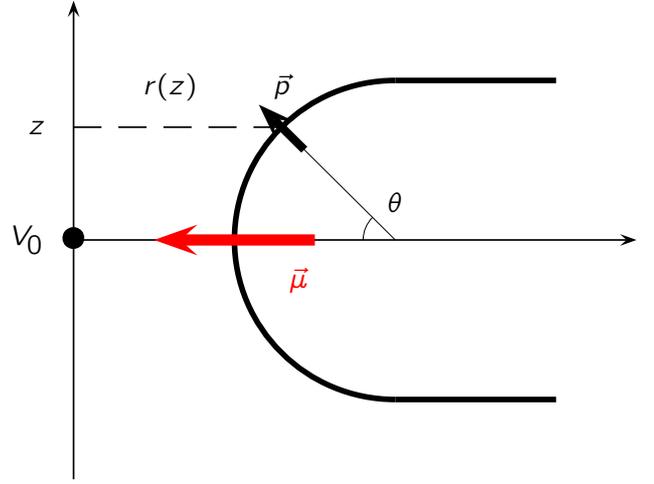}
\figcaption{The effective dipole, $\vec\mu$, 
creates at the origin the same potential, $V_0$, as the molecular 
dipoles, $\vec{p}$, of a slice of angle $\d\f$.}
\label{fig:hole3}
\end{figure}

\begin{figure}
\centering
\includegraphics[width=.9\linewidth]{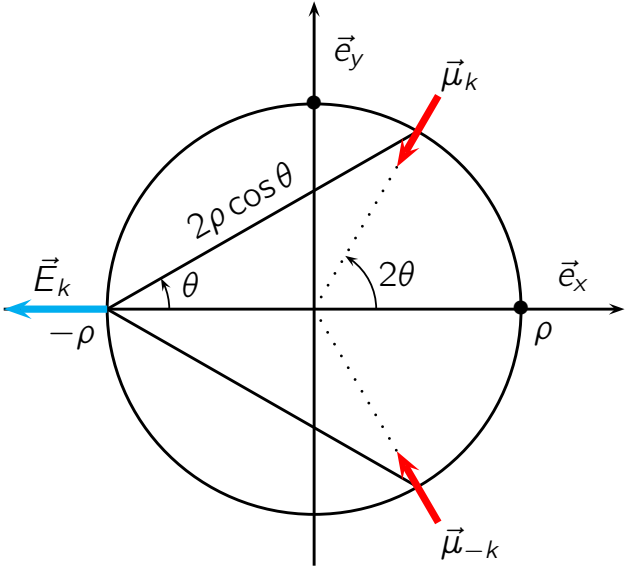}
\figcaption{Each pair $(\vec\mu_k,\vec\mu_{-k})$ 
of effective dipoles ceates an outgoing electric field at $(-\r,0)$.}
\label{fig:hole4}
\end{figure}

In this section, we suppose that protocells can change their 
topology and connect the two layers of the membrane through a toroidal
hole (figure \ref{fig:hole1}).
Following \cite{BergerGostiaux1988}, the principal 
radii of curvature at $B$ are :
\bea
R_1 &= -BA = -\frac{BB'}{\cos\theta} =
-\frac{\r+\e(1-\cos\theta)}{\cos\theta} <0 \\
R_2 &= BC = \e >0 \\
\eea
with $-{\pi}/{2} \leq \theta \leq {\pi}/{2}$.
The distance to the axis of symmetry is 
\bea
r = \r+\e(1-\cos\theta) = BB' .
\eea
The mean curvature at $B$ is :
\bea
H := \frac{1}{2 R_1} + \frac{1}{2 R_2} 
&= \frac{\r+\e(1-2\cos\theta)}{2\e \big( \r+\e(1-\cos\theta) \big)} 
\quad \text{if } r < \r+\e \\
&= 0 \quad \text{if } r > \r+\e .
\eea
Note that $H$ is discontinuous at $\theta = \pm\pi/2$ ($r=\r+\e$), 
but a merit of this expression is to give an upper bound for 
the elastic energy of the hole. If we used a minimal surface 
$(H=0)$ like a catenoid, of equation $r = \r \cosh(z/\r)$, 
we could not glue it smoothly to the horizontal layers
and the singularity would be stronger.
The area measure at $B$ is $\d S = \e\,\d\theta\,r\,\d\f$
and the elastic (bending) energy of this hole is \cite{Boal2012}:
\bea
\E_{\text{bend}} & := \frac{\kappa}{2} \iint H^2 \, \d S 
= \pi\kappa\e \int_{-\pi/2}^{\pi/2} \d\theta \, r H^2 \\
{} &= \frac{\pi\kappa}{4\e} \int_{-\pi/2}^{\pi/2} \d\theta \,
\frac{(r-\e\cos\theta)^2}{r} \\
&= \frac{\pi\kappa}{4\e} \int_{-\pi/2}^{\pi/2} \d\theta 
\left( r-2\e\cos\theta + \frac{\e^2\cos^2\theta}{r} \right)
\eea
where $\kappa$ is the bending modulus of a single layer.
By differentiating under the integral, we obtain the radial 
force density due to this bending energy :
\bea
f_{\r,\text{bend}} &:= -\frac{\p\E_{\text{bend}}}{\p\r} \\
&= -\frac{\pi^2\kappa}{4\e} 
+ \frac{\pi\e\kappa}{4} \int_{-\pi/2}^{\pi/2}
\left( 
\frac{\cos\theta}{\r+\e(1-\cos\theta)} 
\right)^2 \,\d\theta . 
\eea
The asymptotics of $f_\r$, when $\r\to +\infty$, is :
\bea
f_{\r,\text{bend}} & 
= -\frac{\pi^2\kappa}{4\e} + \mathcal{O}(\r^{-2}) < 0 .
\label{fbend1}
\eea
Since $f_{\r,\text{bend}} < 0$ when $\r \gg \e$, large holes 
tend to collapse. However, for small holes $(\r < \e)$ we have :
\bea
f_{\r,\text{bend}} 
&= \frac{\pi^2\kappa\e}{8\r^2} + \mathcal{O}(\r^{-1}) > 0
\qquad (\r \to 0^+) .
\eea
Therefore, small holes avoid total collpase due to the repulsive 
bending forces between the polar heads of the membrane molecules. 
Moreover, for small radius, their electric dipoles repell one another
and cations can form a positively charged layer that stabilises 
the hole, due to electrostatic repulsion at short distances
(figure \ref{fig:hole3}).
\begin{figure}
\centering
\includegraphics[width=.9\linewidth]{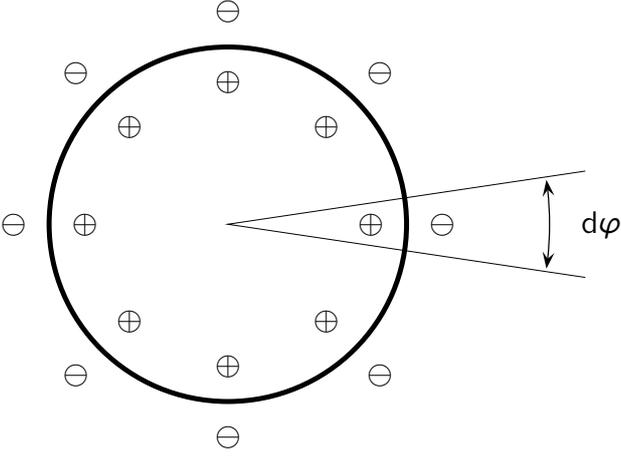}
\figcaption{Equatorial section of a small toroidal hole : 
the repulsive interaction of polar heads forbids the hole to collapse.}
\label{fig:hole2}
\end{figure}
The electrostatic potential created by these dipoles at the center 
of the hole is proportional to $\r^{-2}$ and repulsive, in the dipolar 
approximation. Let us evaluate the effective dipole moment which would
create, at the center of the hole, the same electric field as a meridian 
(vertical) slice of angle $\d\f$ of the membrane 
(figures \ref{fig:hole2} and \ref{fig:hole3}). 
If $\vec{p}$ denotes the dipolar moment of each molecule, occupying 
the area $a$, then a piece of surface of height 
$\d z = \e\,\cos\theta\,\d\theta$, of area 
\bea
\d S &= \e\,\d\theta\, r(z)\,\d\f 
= \frac{r(z)\,\d z\,\d\f}{\cos\theta} \\
\eea
located at the distance $r(z)$ from the axis $\{r=0\}$, 
contains $\d S/a$ parallel dipoles $\vec p$ and
creates at the origin the electric potential
\bea
\d^2 V_0 &= 
\frac{1}{4\pi\e_w}\frac pa \frac{\d S}{r^2(z) + z^2} \\ 
&= \frac{1}{4\pi\e_w}\frac pa 
\frac{r(z)\,\d z\,\d\f}{(r^2(z) + z^2)\cos\theta} ,
\eea
where $\e_w$ denotes the dielectric permittivity of water.
Since $z=\e\sin\theta$ and $r(z)=\r+\e(1-\cos\theta)$, we have :
\bea
r(z) &= \r+\e(1-\sqrt{1-z^2/\e^2}) .
\eea
The total electric potential created at the origin by the piece of
angle $\d\f$ is then :
\bea
\d V_0 &= \frac{\mu}{4\pi\e_w \r^2} \\ 
\mu &:= \frac{p\r^2\e}{a} \int_{-\e}^{\e}
\frac{r(z)\,\d z}{(r^2(z)+z^2)\sqrt{1-z^2/\e^2}}
\eea
The latter, cumbersome expression of $\mu$ is not important.
This is just a number which depends on the exact shape of the surface. 
In order to estimate the electrostatic energy of the hole, 
we can replace the membrane by 
$N = \lfloor {2\pi\r}/{\sqrt a} \rfloor$
regularly spaced dipoles $\vec{\mu}$ on the circle of radius $\r$,
each of these dipoles being directed toward the origin.
Each effective dipole occupies an arc of length $\simeq \sqrt a$
(figure \ref{fig:hole4}).
Using the basis 
\bea
\vec{u} &= -\cos\theta \vec{e}_x -\sin\theta \vec{e}_y \\
\vec{v} &= \sin\theta \vec{e}_x -\cos\theta \vec{e}_y
\eea
the electric field, $\vec{E}_k$, created at $(-\r,0)$ by the $k$-th 
pair of dipoles, $\vec\mu_k$ located at 
$(\r\cos 2\theta,\r\sin 2\theta)$ and $\vec\mu_{-k}$ located at
$(\r\cos 2\theta,-\r\sin 2\theta)$, is :
\bea
\vec{E}_k &= 
\frac{\mu \big( (3\cos^2\theta -1) \,\vec{u} 
- 3\cos\theta \sin\theta \,\vec{v} \big)}{4\pi\e_w (2\r\cos\theta)^3} \\
&= \frac{\mu \big( (3\cos^2\theta -1) \cos\theta 
- 3\cos\theta \sin^2\theta \big)}{4\pi\e_w (2\r\cos\theta)^3}
\, \vec{e}_x \\
&= \frac{\mu}{16\pi\e_w\r^3} 
\left( \frac{1}{\cos^2\theta} -3 \right) \, \vec{e}_x .
\eea
Since $\vec\mu_k = \vec\mu_{-k}$ when $\theta=2k\pi/N=0$, 
the field $\vec{E}_0$ is :
\bea
\vec{E}_0 &= \frac{-\mu}{16\pi\e_w\r^3} \vec{e}_x .
\eea
We can suppose that $N$ is even, for simplicity.
The electrostatic energy of the dipole 
$\vec{\mu}_{N/2} = \mu\,\vec{e}_x$,
located at $(-\r,0)$, is then :
\bea
\E_{\text{dip}} &= -\sum_{k=0}^{\frac N2 -1} \vec\mu \cdot \vec{E}_k \\
&= \frac{\mu^2}{16\pi\e_w \r^3} 
\left( 1 + 3 \left( \frac N2 -1 \right) 
- \sum_{k=1}^{\frac N2 -1} \frac{1}{\cos^2(k\pi/N)}
\right)
\eea
If $N$ is sufficiently large, the trigonometric sum is 
\bea
\sum_{k=1}^{\frac N2 -1} \frac{1}{\cos^2(k\pi/N)}
&\simeq \frac{N^2}{6} 
\eea
and the electrostatic energy becomes :
\bea
\E_{\text{dip}} & \simeq \frac{\mu^2}{16\pi\e_w \r^3} 
\left( \frac{3N}{2} - 2 - \frac{N^2}{6} \right) \\
&= \frac{\mu^2}{16\pi\e_w} 
\left( \frac{3\pi}{\sqrt a} \r^{-2}
- 2 \r^{-3}
- \frac{2\pi^2}{3a} \r^{-1} \right)
\eea
If we neglect the variations of $\mu$ with $\r$, we obtain the
dipolar force exerted by the hole on each dipole $\vec{\mu}$ :
\bea
f_{\r,\text{dip}} &:= -\frac{\p\E_{\text{dip}}}{\p\r} \\
&= \frac{-\mu^2}{16\pi\e_w} 
\left( \frac{-6\pi}{\sqrt a} \r^{-3}
+ 6 \r^{-4}
+ \frac{2\pi^2}{3a} \r^{-2} \right) .
\eea
For large holes, these three terms are dominated by the constant 
cohesion force, $f_{\r,\text{bend}}(\infty) = -\pi^2 \kappa/4\e$ 
(Eq. \ref{fbend1}).
However, for small holes $(\r \sim \e)$ the dipolar approximation 
becomes dubious. The incompressibility of membrane molecules enters
the stage and the electrostatic repulsive forces between their polar
heads avoid the collapse of the hole. Moreover, the large $N$
approximation is not valid anymore for small $\r$.
These estimates of $f_{\r,\text{dip}}$ and $f_{\r,\text{bend}}$ 
suggest that bare nanoholes (without intrinsic proteins) 
reach a (meta)stable equilibrium between attractive elastic forces 
($\r$ large) and repulsive electrostatic forces ($\r$ small).
The energy cost of a toroidal hole of radius 
$\r \sim \e \sim 10^{-9}$ m, is :
\bea
\E & \simeq \frac{\pi^2 \kappa}{4} 
\simeq 2.5\, \kappa
\simeq \kappa_b
\simeq 10^{-19} \, J
\simeq 25 \, k_B T
\eea
according to \cite{Boal2012}. Multiplying by the Avogadro 
number, we obtain the molar energy of holes $\simeq 10$ kJ/mol.
This value is, however, an overestimated upper bound, since the hole
tends to minimise its bending energy and adopt 
a shape between a torus and a catenoid ($H=0$). 
Therefore, we conjecture that the true minimum of $\E$ is 
smaller than $10$ kJ/mol. 
Once this price has been paid, many membrane molecules 
can easily flow from the inner leaflet to the outer leaflet, under 
the effect of larger thermal fluctuations inside than outside. 
In this translocation process, the hydrophilic polar heads
do not need to plunge inside the hydrophobic layer 
\cite{Allhusen2017,Anglin2010,Contreras2010,GV1,GV2}
since they bypass this barrier by going through the hole.

\section{\bf Translocation current through holes}

\begin{figure}
\centering
\includegraphics[width=.8\linewidth]{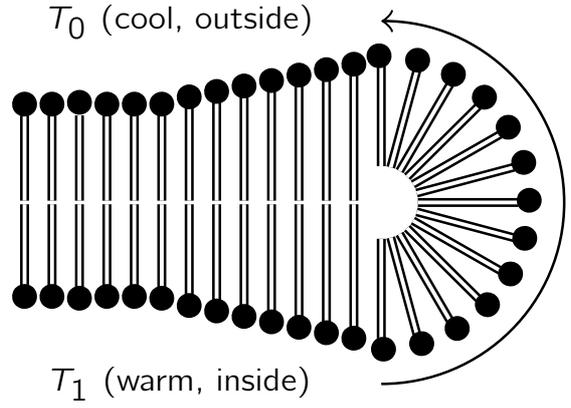}
\figcaption{Radial section of the membrane near a hole. 
An outgoing molecular current results from 
the temperature gap, $T_1-T_0 > 0$.}
\label{fig:hole5}
\end{figure}

In this section, we compute an estimate of the current of 
membrane molecules through a hole (figure \ref{fig:hole5}).
This current is induced by the temperature difference between the
inside and the outside of the protocell. A single hole allows 
the membrane molecules to bypass the large energy barrier of 
translocation through the median hydrophobic layer \cite{JC}.
The flow of molecules from the inner leaflet to the outer leaflet
is, however, always kinetically limited by the height, $E_\ast$, 
of the energy barrier due to the bending energy and to the electrostatic 
potential energy, $V$ (we take $V_{\theta=\pm\pi/2}=0$). If $a$ denotes 
the average area occupied by each polar head, then we have :
\bea
E_\ast &= \max \left(\frac{\kappa}{2} H^2 a + V \right) .
\eea
This maximum is reached at the equator, $\{\theta = 0\}$ :
\bea
E_\ast &= 
\frac{\kappa a}{8} \left( \frac{1}{\e} - \frac{1}{\r} \right)^2 
+ V_{\theta = 0} . \\ 
\eea

Let us give a rough estimate of the net flow of membrane molecules
through a hole.
The average number of molecules per unit area in each leaflet is $1/a$. 
If $\bar{v}_{h1}$ denotes the average velocity of the particles
which are quick enough to cross the energy barrier, $E_\ast$, and fall
into the hole, of radius $\r$, then the flow of molecules leaving 
the inner leaflet and reaching the outer leaflet is :
\bea
j_{m1} &= \frac{2\pi\r \bar{v}_{h1}}{a} .
\eea
As in \cite{TDFP}, we use an ideal gas model to evaluate 
$\bar{v}_{h1}$. Let us introduce the following notations : 
\bea
v_\ast &= \sqrt{2E_\ast/m} = \sqrt{2 k_B T_\ast /m} \\
\bar{v}_1 &= \sqrt{2 k_B T_1/m} .
\eea
The average velocity, $\bar{v}_{h1}$, of the molecules that 
succeed in crossing the barrier is obtained by integrating 
the Maxwell-Boltzmann law over the half-line of sufficient 
incoming radial velocities :
\bea
\bar{v}_{h1} &= \frac{1}{2\bar{v}_1 \sqrt\pi} 
\int_{v_\ast}^{\infty} \d v\, v \, \exp(- v^2/\bar{v}_1^2) \\
&= \frac{\bar{v}_1}{4\sqrt\pi} \int_{v_\ast^2/\bar{v}_1^2}^{\infty}
\d s \, e^{-s} 
= \frac{e^{- v_\ast^2/\bar{v}_1^2}}{4\sqrt\pi}
\frac{\bar{v}_1^3}{v_\ast^2} .
\eea
Therefore, we have : 
\bea
j_{m1} &= \frac{\sqrt\pi\r}{2a} \, 
\frac{\bar{v}_1^3}{v_\ast^2} \, 
e^{- v_\ast^2/\bar{v}_1^2} \\
j_m &:= j_{m1} - j_{m0} 
= \frac{\sqrt\pi \r }{2 a v_\ast^2}
\left(
\bar{v}_1^3 e^{- v_\ast^2/\bar{v}_1^2}
- \bar{v}_0^3 e^{- v_\ast^2/\bar{v}_0^2}
\right) \\
&= \sqrt{\frac{\pi k_B}{2m}} \frac{\r}{a T_\ast}
\left(
T_1^{3/2} e^{- T_\ast/T_1}
- T_0^{3/2} e^{- T_\ast/T_0}
\right) . \\
\eea
If $T_1 > T_0$ then $j_m > 0$ : we obtain an outgoing molecular 
flow. A more precise model would change the geometric form factor
$\sqrt{\pi/2}$ and would provide corrections due to molecular
interactions (surface tension), but it would preserve 
the leading term $(T^{3/2}/T_\ast)$ and the exponential factor.

The crossed conductance coefficient of a single hole, $\ell_{m\theta}$,
is defined as the derivative of $j_m$ with respect to $1/T_1$ :
\bea
\ell_{m\theta} &:= \frac{\p j_m}{\p (1/T_1)} 
= - \sqrt{\frac{\pi k_B}{2m}} \frac{\r}{a T_\ast}
T_1^{3/2} e^{- T_\ast/T_1}
\left( \frac 32 T_1 + T_\ast \right)
\eea
$\ell_{m\theta}$ is negative because the molecular current decreases
when the inverse temperature, $1/T_1$, increases (the inside 
cools down).

\section{\bf Conclusion and perspectives}

The possibility that protocells be bounded by a punctured bilayer 
is based on the existence of transient or stable nanoholes. 
The lifetime of the latter depends also on the molecular surrounding 
of the membrane.
In particular, the properties of water in the nanometric domain are not 
the same as in the macroscopic scale : the Reynolds number is very low, 
flows are laminar and surface tension is a dominant force. 
Moreover, in a non-equilibrium setting, the emission of infrared radiation
from the interior of the protocell enhances the formation 
of an exclusion zone \cite{Pollack2013,Elton2020} that could give nanoholes 
unusual properties. Moreover, such a two-dimensional system opens the door 
to quantum coherence effects \cite{Sen2015} that could stabilise
these nanoholes.

If life starts with cellularity and the ability 
to split \cite{Mor2}, induced by some inner, sufficiently exothermic 
metabolism, then a change of topology as described above facilitates 
drastically the translocation of membrane molecules toward 
the outer layer, until a complete splitting of the vesicle.
In a sense, topology change could be at the basis of life itself,
by catalysing the splitting process proposed in \cite{TDFP}, 
but only experiments will confirm or reject our hypothesis.

\bibliographystyle{biophysj}
\bibliography{top}



\end{document}